\documentstyle[12pt]{article}    
\begin{document}           
\baselineskip=0.33333in
\begin{quote} \raggedleft TAUP 2818-2005
\end{quote}
\vglue 0.5in
\begin{center}{\bf Theoretical Errors in\\
Contemporary Physics}
\end{center} 
\begin{center}E. Comay 
\end{center}
 
\begin{center}
School of Physics and Astronomy \\
Raymond and Beverly Sackler Faculty of Exact Sciences \\
Tel Aviv University \\
Tel Aviv 69978 \\
Israel
\end{center}

Email: elic@tauphy.tau.ac.il
\vglue 0.5in
\vglue 0.2in

\vglue 0.2in
\noindent
Abstract:  Errors pertaining to the following physical theories are
discussed: the Dirac magnetic
monopole theory; the Klein-Gordon equation; the Yukawa
theory of nuclear force; the idea of Vector Meson Dominance; the
Aharonov-Bohm effects; the idea of diffraction-free electromagnetic
beams; Quantum Chromodynamics. Implications of the theoretical errors
are discussed briefly. In particular, relations between the Dirac
monopole theory, the idea of Vector Meson Dominance and Quantum
Chromodynamics cast doubt on the current interpretation of strong
interactions.

\newpage
\noindent
{\bf 1. Introduction}
\vglue 0.33333in

   The purpose of the present work is to discuss several theoretical
errors existing in contemporary physics. Before addressing specific cases,
let us examine the structure of a physical theory and the meaning of
errors that can be found in it.

   A physical theory resembles a mathematical theory. Both rely on a set
of axioms and employ a deductive procedure for yielding theorems, corollaries
etc. The set of axioms and their results are regarded as elements of the 
structure of the theory.
However, unlike a mathematical theory, one requires that a physical
theory should explain existing experimental data and predict results of
new experiments.

   This distinction between a mathematical theory and a physical theory
has several aspects. First, experiments 
generally do not yield precise values but
contain estimates of the associated errors. (Some quantum mechanical
data, like spin, make an exception.) It follows that in many cases,
a certain numerical difference between theoretical predictions and
experimental data is quite acceptable.

   Next, one does not expect that a physical theory should explain every
phenomenon. For example, it is well known that physical theories yield
very good predictions for the motion of planets around the sun. On the
other hand, nobody expects that a physical theory 
be able to predict the
specific motion of an eagle flying in the sky. This simple example proves
that the validity of a physical theory should be evaluated only with
respected to a limited set of experiments. The set of experiments which
are relevant to a physical theory is called its domain of validity. (A
good discussion of this issue can be found in [1], pp. 1-6.)

   Relations between two physical theories can be deduced from an
examination of their domain of validity. In particular, let $D_A$ and
$D_B$ denote the domains of validity of theories $A$ and $B$,
respectively. Now, if $D_A \subset D_B$ and $D_A \neq D_B$ then
one finds the theory
$B$ takes a higher hierarchical rank with respect to theory $A$ (see
[1], pp. 3-6). Here theory $B$ is regarded as a theory having a more
profound status. However, theory $A$ is not ``wrong", because it yields
good predictions for experiments belonging to its own (smaller) domain
of validity. Generally, theory
$A$ takes a simpler mathematical form. Hence,
wherever possible, it is used in actual calculations. 
Moreover, since theory $A$ is good in its validity domain
$D_A$ and $D_A \subset D_B$ then one finds that {\em theory $A$ imposes
constraints on theory 
$B$, in spite of the fact that $B$'s rank is higher than
$A$'s rank}. This self evident relation between a lower rank theory and
a higher one is called below ``restrictions imposed by a lower rank
theory". It is used here more than once.
Thus, for example, although Newtonian
mechanics is good only for cases where the velocity $v$ satisfies
$v\ll c$, relativistic mechanics should yield formulas 
which agree with corresponding formulas of Newtonian mechanics, 
provided $v$ is small enough.

   Having these ideas in mind, a theoretical error is regarded here as a
mathematical part of a theory that yields predictions which are clearly
inconsistent with experimental results, where the latter are carried out
within the theory's validity domain. The direct meaning of this definition
is obvious. It has, however, an indirect aspect too. Assume that
a given theory has a certain part, $P$, which is regarded as well
established. Thus, let $Q$ denote
another set of axioms and formulas which yield
predictions that are inconsistent with $P$. 
In such a case, $Q$ is regarded as a
theoretical error. (Note that, as explained above, $P$ may belong to a lower
rank theory.) An error in the latter sense is analogous to an
error in mathematics, where two elements of a theory are inconsistent
with each other.

   There are other aspects of a physical theory which have
a certain value but are not well defined.
They may be described as neatness, simplicity and physical acceptability of
the theory. A general rule considers theory $C$ as simpler (or neater)
than theory $D$ if theory $C$ relies on a smaller number of axioms.
These properties of a physical theory are relevant to a
theory whose status is still undetermined because there is a lack of
experimental data required for its acceptance or rejection.

   The notions of neatness, simplicity and physical acceptability have
a subjective nature and it is not clear how a disagreement based on them
can be settled. In particular, one should note that ideas concerning
physical acceptability changed dramatically during the 20th century. Thus,
a physicist of the 19th century would have regarded many well established
elements of contemporary physics as unphysical. An incomplete list of
such elements contains the relativity of length and time intervals, the
non-Euclidean structure of
space-time, the corpuscular-wave nature of pointlike particles,
parity violation and the nonlocal nature of quantum mechanics (which is
manifested by the EPR effect).

   For these reasons, neatness, simplicity and physical acceptability
of a theory have a secondary value. Thus, if there is no further evidence
then they should not be used for taking a {\em final decision} concerning the
acceptability of a physical theory. In this work, properties of a
physical theory pertaining to a lack of neatness, simplicity and physical
acceptability are mentioned. 
However, this aspect of the problems may be helpful for
the reader but they should not be regarded as decisive arguments. In the text
there is no distinction between neatness and simplicity. Thus, the term
neatness is not used.

   Before concluding these introductory remarks, it should be stated that the
erroneous nature of a physical theory $E$ cannot be established just 
by showing the
existence of a different (or even a contradictory) theory $F$. This point 
is obvious. Indeed, if such a situation exists then one may conclude that 
(at least) theory $E$
{\em or} theory $F$ is wrong. However, assuming that neither $E$ nor $F$ rely on
a mathematical
error, then one cannot decide on this issue without having an adequate
amount of experimental data.

   Another issue is the usage of models and phenomenological formulas. 
This approach
is very common in cases where there is no good theory or where 
theoretical formulas are
too complicated. This approach is evaluated by its usefulness and not by its
theoretical correctness. Hence, it is not discussed in the present work.

   The following discussions rely on the ideas 
described above and are devoted to theoretical aspects of the
following topics: the Dirac magnetic monopole 
(called just monopole) theory, the Klein-Gordon
(KG) equation, the Yukawa interaction, the idea of Vector Meson Dominance
(VMD), the Aharonov-Bohm (AB) effects and the idea of creating
diffraction-free electromagnetic beams. Experimental data pertaining
to Quantum Chromodynamics that have no adequate explanation are
presented in the penultimate Section. The paper contains new material
that has not been published yet and other topics that have already been
published. The latter cases are included here in order to help the
reader see the full picture. However, the corresponding
presentation takes a concise form and references to
detailed articles are given.

   In this work units where $\hbar = c = 1$ are used. The Lorentz metric is
diagonal and its entries are $(1,-1,-1,-1)$. Greek indices run from 0 to 3.
The symbol $W_{,\mu }$ denotes the partial derivative of $W$ with
respect to $x^\mu $.

\vglue 0.66666in
\noindent
{\bf 2. The Dirac Monopole Theory}
\vglue 0.33333in

   Monopoles are defined by the following duality transformation (called
also duality rotation by $\pi /2$)
\begin{equation}
\bf E \rightarrow \bf B,\;\;\;\bf B \rightarrow -\bf E
\label{eq:EB}
\end{equation}
and
\begin{equation}
e \rightarrow g,\;\;\;g\rightarrow -e,
\label{eq:EG}
\end{equation}
where $g$ denotes the magnetic charge of monopoles.

   A theory of monopoles was published by Dirac in the first half of the
previous century[2,3]. At present, there is no established experimental
evidence of these monopoles[4]. This experimental status of monopoles
led Dirac later in his life to state: ``I am inclined now to believe that
monopoles do not exist. So many years have gone by without any
encouragement from the experimental side" [5].

   Here the following question arises: does the failure of the 
monopole quest stems from the fact that they do not exist 
in Nature or from erroneous
elements in the Dirac's monopole theory? It is shown in this Section that the
second possibility holds.

   Let us examine the established part of electrodynamics. Here the system
consists of electric charges carried by matter particles and
electromagnetic fields. The equations of motion of the fields are
Maxwell equations
\begin{equation}
F^{\mu \nu}_{(e)\;,\nu} = -4\pi j_{(e)}^\mu
\label{eq:MAX1}
\end{equation}
and
\begin{equation}
F^{*\mu \nu}_{(e)\;,\nu} = 0.
\label{eq:MAX2}
\end{equation}
and the 4-force exerted on charged matter is given by the Lorentz law
\begin{equation}
ma_{(e)}^\mu = eF_{(e)}^{\mu \nu }v_\nu.
\label{eq:LOR}
\end{equation}
Here $F^{\mu \nu }$ is the antisymmetric tensor of the electromagnetic fields, 
$F^{*\mu \nu } = \frac {1}{2} \varepsilon ^{\mu \nu \alpha \beta } F_{\alpha \beta}$,
$\varepsilon ^{\mu \nu \alpha \beta}$ is the completely antisymmetric
unit tensor of the fourth rank and $j^\mu $ is the electric 4-current.
Subscripts $_{(e)},\;_{(m)}$ denote quantities related to charges and
monopoles, respectively. The duality transformation of fields $(\!\!~\ref{eq:EB})$
can be written in a tensorial form $F^{\mu \nu} \rightarrow F^{*\mu \nu}$.

   An important quantity is the electromagnetic 4-potential $A_\mu $. This
quantity is used in the Lagrangian density of the system. The fields'
part of the Lagrangian density is (see [6], p. 71; [7], p. 596)
\begin{equation}
L_{fields} = -\frac {1}{16\pi }F_{(e)}^{\mu \nu }F_{(e)\mu \nu } -j_{(e)}^\mu A_{(e)\mu}.
\label{eq:LINT}
\end{equation}

   Using the duality transformation $(\!\!~\ref{eq:EB})$, 
$(\!\!~\ref{eq:EG})$ and Maxwellian electrodynamics
$(\!\!~\ref{eq:MAX1})$-$(\!\!~\ref{eq:LINT})$, 
one derives a dual Maxwellian theory for a system of monopoles
and electromagnetic fields (namely, a system without charges)
\begin{equation}
F^{*\mu \nu}_{(m)\;,\nu} = -4\pi j_{(m)}^\mu,
\label{eq:MAX1M}
\end{equation}

\begin{equation}
-F^{\mu \nu}_{(m)\;,\nu} = 0.
\label{eq:MAX2M}
\end{equation}

\begin{equation}
ma_{(m)}^\mu = gF_{(m)}^{*\mu \nu }v_\nu.
\label{eq:LORM}
\end{equation}
and
\begin{equation}
L_{fields} =  -\frac {1}{16\pi }F_{(m)}^{*\mu \nu }F^*_{(m)\mu \nu }  -j_{(m)}^\mu A_{(m)\mu}.
\label{eq:LINTM}
\end{equation}

   At this point we have two theories: the ordinary Maxwellian
electrodynamics whose domain of validity does not contain 
magnetic monopoles and
a monopole related Maxwellian theory which does not contain electric charges.
The problem is to determine the form of a covering theory of a system of
charges, monopoles and their fields.

   As explained in the first Section, the two subtheories mentioned above
impose constraints on the required charge-monopole theory:

\begin{itemize}
\item[{1.}] It should conform to Maxwellian electrodynamics 
$(\!\!~\ref{eq:MAX1})$-$(\!\!~\ref{eq:LINT})$ in the limit
where monopoles do not exist.

\item[{2.}] It should conform to the dual Maxwellian electrodynamics 
$(\!\!~\ref{eq:MAX1M})$-$(\!\!~\ref{eq:LINTM})$
in the limit where charges do not exist.
\end{itemize}

   It turns out that Dirac's monopole theory is inconsistent with requirement
2. Therefore, it is inconsistent with a restriction imposed by a lower
rank theory.

   As a matter of fact, Dirac also uses implicitly a new axiom which has
no experimental support. Thus, his theory assumes that:

\begin{itemize}
\item[{A.}]
Electromagnetic fields of charges and electromagnetic fields of
monopoles have identical dynamical properties.
\end{itemize}

This approach forces him to use just one kind of 4-potential $A_\mu $ and
to confront a new kind of singularity. Indeed, if the 3-vector $\bf A$ 
is regular then
\begin{equation}
\nabla \cdot \bf B = \nabla \cdot (\nabla \times \bf A) = 0
\label{eq:DIVB}
\end{equation}
and monopoles do not exist. Dirac uses the term 'string' for this kind
of singularity.  The utilization of the new axiom A, and the introduction
of a new kind of singularity into electrodynamics indicate a departure
from simplicity.

   Several additional errors of the Dirac monopole theory have been pointed out
a long time ago. Thus, it was claimed that the Dirac monopole theory
is inconsistent with the S-matrix theory (see [8,9]). A third article [10]
claims that the inclusion of the Dirac monopole in electrodynamics is
inconsistent with relativistic covariance. Another kind of error of the
Dirac monopole theory was published recently [11]. It is sown there
that a hypothetical quantum mechanical system that contains a charge and
a Dirac monopole violates energy conservation (see [11] pp. 98-99).

   Another  problem is the definition of the interaction part of
the angular momentum in a system containing an electric charge and a Dirac
monopole. Here one finds that the interaction part of the fields'
angular momentum {\em does not vanish for cases where
the distance between the two particles
tends to infinity} (see [7] p. 256; [11], pp. 97-98; [12] p. 1366).
Such a kind of interaction is unknown in classical electrodynamics and is
regarded as unphysical.

   The discussion carried out in this Section shows 
several theoretical errors and a deviation from simplicity done by using
an additional axiom and unphysical properties of the Dirac monopole theory.
These difficulties are completely consistent
with the failure of the experimental efforts aiming to detect Dirac monopoles.
It is interesting to note that a regular and self-consistent
charge-monopole theory can be
constructed without using axiom A [11,13,14]. This theory derives a
different set of equations of motion. The failure of the attempts to detect
the Dirac monopoles is predicted in [8] and it  
is derived from the equations of motion of the
regular monopole theory [15] as well.

\vglue 0.66666in
\noindent
{\bf 3. The Klein-Gordon Equation}
\vglue 0.33333in

   The KG equation
\begin{equation}
(\Box + m^2)\phi = 0
\label{eq:KGEQ}
\end{equation}
was derived in the very early days of quantum mechanics (see [16], bottom
of p. 25). It
can be regarded as a quantize form of the relativistic relation
$E^2 - {\bf p}^2 = m^2$, where $i\partial/\partial t,\; -i\nabla$ replace 
$E$ and
$\bf p$, respectively. Hence, there is no doubt concerning its correctness
{\em as
a formula}. Indeed, as is well known, components of a solution of the Dirac
equation satisfy the KG equation.

   The problem discussed in this Section is the status of the KG equation
$(\!\!~\ref{eq:KGEQ})$ {\em as a fundamental quantum mechanical equation 
derived from a
Lagrangian density}. Here the Lagrangian density of an electrically charged 
KG particle is
\begin{equation}
{\mathcal L} = (\phi ^*_{,0} -ieV\phi ^*)(\phi _{,0}+ ieV\phi) -
\sum _{k=1}^3 (\phi _{,k}^* +ieA_k \phi ^*)(\phi _{,k} -ieA_k \phi )
- m^2\phi ^* \phi.
\label{eq:PWLD}
\end{equation}
(See [17,18], eq. (37). Note that here units where $\hbar = c=1$ are introduced.) This
aspect of the KG equation took a controversial status for a very long 
time. Dirac's
negative opinion on this equation (see [19] and [20], pp. 3-8) directed him 
to construct his
famous equation which is now regarded as the relativistic quantum mechanical
Hamiltonian of spin-1/2 particles.

   Other researchers disagree with Dirac (see [18], pp. 70-72, 105, 
188-205; [16],
second column of p. 24). In particular, Pauli and Weisskopf constructed 
the second
order Lagrangian density $(\!\!~\ref{eq:PWLD})$.  
Unlike the case of the Dirac equation, this Lagrangian density 
does not
yield an expression for the particle density but for its charge density.

    Before examining the experimental side, let us state a fundamental 
property of
particles described by a wave function $\psi (x^\mu)$. Due to the fact 
that $\psi (x^\mu)$
depends on a {\em single} set of space-time coordinates $x^\mu$, one 
concludes that
a particle {\em truly described} by $\psi (x^\mu)$ must be elementary, namely 
a pointlike structureless particle.

   The experimental data of elementary massive spin 1/2
(Dirac) particles, like the electron, the muon and the
$u,d$ quarks is consistent with the pointlike requirement. This is not true
for the
old candidates for the KG particles, namely the three $0^-$ $\pi $ 
mesons. Indeed,
it is now known that a $\pi $ meson contains a quark and an antiquark.
The charge radius of the $\pi ^\pm $ is $0.672 \pm 0.008$ fm 
(see [4], p.499). Hence, 
$\pi $ mesons are definitely not pointlike particles.

   A recent analysis of the KG Lagrangian density proves that it is also 
not free of
theoretical difficulties [21]. Thus, it is proved that the theory derived 
from the KG
Lagrangian density $(\!\!~\ref{eq:PWLD})$ has the following difficulties:
\begin{itemize}
\item[{1.}] There is no expression for the particle's density. The expression for the
charge density depends on coordinates of {\em external particles}.
\item[{2.}] The Hamiltonian density depends on time derivative of $\phi $. Hence,
if a Hamiltonian of the KG particle exists then the Hamiltonian density 
depends on the Hamiltonian.
\item[{3.}] There is no {\em covariant differential operator} that serves as 
a Hamiltonian [21].
Furthermore, the Hamiltonian matrix of a charged KG particle destroys the 
inner product of
the Hilbert space [21]. There is no Hilbert space for an uncharged KG 
particle because in this case density is undefined [17].
\item[{4.}] The {\em second order} KG equation $(\!\!~\ref{eq:KGEQ})$, 
which is derived from
the KG Lagrangian density $(\!\!~\ref{eq:PWLD})$, is not identical to 
the {\em first order}
fundamental quantum mechanical equation $i\partial \phi /\partial t = H\phi $.
\item[{5.}] One cannot construct a self-consistent electromagnetic 
interaction of a
charged KG particle. The linear interaction $eA_\mu j^\mu$ entails an 
equation imbalance [22]
and the quadratic term $(p^\mu -eA^\mu)(p_\mu - eA_\mu)$ destroys the 
inner product of the Hilbert space [21].
\item[{6.}] There is no explanation why the energy-momentum operators $(i\partial /\partial t,
-i\nabla)$ are used for the {\em different} task of representing charge density and current.
\item[{7.}] The nonrelativistic limit of the KG equation disagrees with the
Schroedinger equation. Indeed, in the case of the Schroedinger equation,
$\Psi^*\Psi$ represents probability density [23] whereas the KG
equation has no expression for probability density. Hence, the KG
equation is inconsistent with a restriction imposed by a lower rank
theory. 
\end{itemize}

   (By contrast, it is proved in [21] that an analogous analysis of the Dirac equation 
yields completely acceptable relations.)

   These theoretical difficulties, together with the lack of support from the experimental
side (there is no candidate for a {\em pointlike} KG particle)
indicate that, unlike the case of the Dirac equation, the existence of a genuine KG
particle is not very likely.

\vglue 0.66666in
\noindent
{\bf 4. The Yukawa Interaction}
\vglue 0.33333in

   The Yukawa interaction is derived from the interaction term of a Dirac
spinor with a KG particle (see [24], p.79 and [25], p. 135)
\begin{equation}
L_{Yukawa} = L_{Dirac} + L_{KG} - g\bar {\psi}\psi \phi .
\label{eq:LYUKAWA}
\end{equation}
Here the KG particle plays a role which is analogous to that of the 
photon in electrodynamics.
The dependence of $(\!\!~\ref{eq:LYUKAWA})$ on the KG Lagrangian density indicates that
it suffers from all the difficulties of the KG theory which are pointed out
in the previous Section. 
Furthermore, note that, due to the fact that all terms of the Lagrangian
density are Lorentz scalars, the interaction term of 
$(\!\!~\ref{eq:LYUKAWA})$ depends on
the Dirac particle's {\em scalar density}
$\bar {\psi}\psi $ which is {\em not} its 
actual density $\psi ^\dagger \psi $.
This situation is very strange because 
one expects that the intensity of the interaction 
of a Dirac particle should 
depend on its actual density $\psi ^\dagger \psi $ which is a component
of the Dirac 4-current and {\em not} on 
the scalar density $\bar {\psi}\psi $. Moreover, 
it is explained below that $(\!\!~\ref{eq:LYUKAWA})$
is not free of covariance problems.

   An analysis of the nonrelativistic limit of two Dirac particles 
interacting by means of a Yukawa field, yields
the following expression for the interaction term (see [26], p. 211)
\begin{equation}
V(r) = g^2\frac {e^{-\mu r}}{r}    
\label{eq:HYUKAWA}
\end{equation}
where $\mu $ denotes the mass of the KG particle. The Yukawa theory was
suggested as a theoretical interpretation of the nucleon-nucleon interaction.
This idea was proposed
in the early days of nuclear theory when 
nucleons were regarded as elementary Dirac
particles. Now it is known that nucleons are composite particles containing
quarks and this application of $(\!\!~\ref{eq:LYUKAWA})$ is deprived of its theoretical
basis. Furthermore, a recent discussion proves that the classical limit
of the interaction $(\!\!~\ref{eq:HYUKAWA})$ is inconsistent with special relativity
(see [22], p. 13). This argument relies on the relativistic
relation between the 4-velocity and the 4-acceleration
\begin{equation}
a^\mu v_\mu = 0. 
\label{eq:VA}
\end{equation}  
Examining an elementary classical particle, one finds that
relation $(\!\!~\ref{eq:VA})$ yields for the 4-force $f^\mu v_\mu = 0$.
It is explained below why this relation
is inconsistent with the Yukawa interaction $(\!\!~\ref{eq:HYUKAWA})$.

   Let an elementary classical particle $W$
move in a field of force. The field quantities are independent of the 
4-velocity of
$W$ but the associated 4-force must be orthogonal to it. In 
electrodynamics this goal
is attained by means of the Lorentz force $(\!\!~\ref{eq:LOR})$. 
In this case, one finds
\begin{equation}
a^\mu v_\mu = \frac {e}{m}F^{\mu \nu}v_\mu v_\nu = 0, 
\label{eq:LOROK}
\end{equation}  
where the null result is obtained from the antisymmetry of 
$F^{\mu \nu }$ and the
symmetry of the product $v_\mu v_\nu $. In electrodynamics, the 
antisymmetric field tensor
$F^{\mu \nu }$ is constructed as the 4-curl of the 4-potential $A_\mu $. Such a
field of force cannot be obtained from the {\em scalar} KG field. Hence, the
classical limit of the Yukawa interaction is inconsistent with special relativity.

Considering the experimental side,
the application of the Yukawa theory to nuclear interactions cannot be
regarded a success.  The nuclear force is characterized by a very hard (repulsive) core
and a rapidly decreasing attractive force outside this core. Therefore,
at a certain point of $r$, the nuclear potential {\em changes sign}
(see [27], p. 97).
The Yukawa formula $(\!\!~\ref{eq:HYUKAWA})$ is inconsistent with this
property. The nuclear force has also a 
tensorial component as well as a spin-orbit dependence (see [27], pp. 68-78).
Today people use phenomenological formulas
for a description of the nucleon-nucleon interaction data (see [27], pp. 97-99).

\vglue 0.66666in
\noindent
{\bf 5. The Idea of Vector Meson Dominance}
\vglue 0.33333in

The idea of VMD has been suggested as an explanation for
interaction properties of high energy photons with hadrons. Here the data
show that the cross section of the interaction of such photons 
with a proton target is very
similar to that of a neutron target [28]. Since the electric charge
of proton constituents differ from those of a neutron, one concludes 
that the interaction of these photons with the 
{\em electric charge} of constituents of
nucleons {\em cannot} explain this similarity.

At first, the VMD idea was not accepted by all physicists. The humoristic-
sarcastic poster published on page 267 of [28] 
provides an illustration for this claim.
Moreover, contemporary classifications of physical subjects (like PACS
and arXiv.org) regard VMD as a {\em phenomenological} idea. Now, if VMD is
just a phenomenological idea or a model
then the current approach of the physical
community to strong and electromagnetic interactions (namely, the
Standard Model) {\em has no theoretical explanation for the photon-hadron
interaction}.

The main idea of VMD is that the wave function of an energetic photon 
takes the form
\begin{equation}
\mid \gamma >\; = c_0\mid \gamma _0> + c_h \mid h>
\label{eq:GAMMA}
\end{equation}
where $\mid \gamma >$ denotes the wave function of a physical photon, 
$\mid \gamma _0>$
denotes the pure electromagnetic 
component of a physical photon and $\mid h>$ denotes 
its hypothetical hadronic component. $c_0$ and $c_h$ are appropriate
numerical coefficients. 
The values of $c_0$ and $c_h$ depend on the photon's energy. Thus, for soft
photons $c_h = 0$ whereas it 
begins to take a nonvanishing value for photons whose
energy is not much less then the $\rho $ meson's mass (see
[28] and [29]).

Theoretical aspects of VMD were discussed recently [30]. This analysis
proves that VMD is inconsistent with well established physical theories
and with experimental data as well. In particular, it is 
proved in [30] that
VMD is inconsistent with Wigner's analysis of the Poincare group [31,32]
and with the scattering data of linearly polarized photons impinging on
an unpolarized target of protons [30].

The following simple thought experiment disproves the VMD's idea 
stating that the size of the
hadronic components of a photon depend on its energy [29]. Consider two
intersecting rays of optical photons (see fig. 1).  In the laboratory
frame $\Sigma $, the optical
photons of the rays do not interact. Thus, neither energy
nor momentum are exchanged between the rays. Therefore, 
after passing through $O$, the photons travel in their
original direction. Let us examine the situation in a frame $\Sigma '$. In
$\Sigma $, frame $\Sigma '$ is seen moving 
very fast in the negative direction of
the Y axis. Thus, in $\Sigma '$, photons of the two rays are very
energetic. Hence, if VMD holds then photons of both rays contain hadrons
and should exchange energy and momentum at point $O$. 
This is a contradiction because if the
rays do not exchange energy and momentum in frame $\Sigma $ then they
obviously do not do that in any other frame of reference. This argument
proves that VMD is a theoretical error.

\vglue 0.66666in
\noindent
{\bf 6. The Aharonov-Bohm Effects}
\vglue 0.33333in

The AB effects refer to the phase difference  
between two sub-beams of an electron
that travels in a non-simply connected field free region [33]. The phase
difference is manifested by the interference pattern of the sub-beams
(see fig. 2). Hereafter, an electron of the beam is called ``the traveling
electron".
The authors of [33] claim that there are two kinds of realization of this
idea. In the electric AB effect, the region $R$ contains a
{\em time-dependent} electric field whereas in the magnetic AB effect
the region $R$ contains a magnetic field.

The AB effects certainly belong to quantum mechanics, because the
sub-beams move in a field free region. Hence, no force is exerted on
the traveling electron and its inertial motion is not affected by the
field at $R$. However, quantum mechanical equations of motion depend on
the 4-potential $A_\mu $. Hence, a quantum mechanical effect may take place.
The effect emerges from the different phase associated with the
sub-beams and is detected by the interference pattern on the screen $S$.
Hence, both the origin and the detection of the effects belong to
the realm of quantum mechanics.

The original approach of the authors of [33] treat the phase as a single
particle property of the traveling electron. This 
approach certainly does not hold in many cases.
Indeed, the quantum mechanical system consists of the traveling electron
{\em and} of the charges associated with the field at $R$. Let $r_e$ and $r_s$
denote the coordinates of the traveling electron and of the charges at
the source of the field, respectively. Thus, since the traveling
electron interacts with the 4-potential $A_\mu $ associated with $r_s$,
one finds that the Hamiltonian of the system takes the form

\begin{equation}
H = H(r_s,r_e)  
\label{eq:H}
\end{equation}
and the Schroedinger equation is

\begin{equation}
i\hbar \frac {\partial}{\partial t} \Psi(r_s,r_e) = H(r_s,r_e)\Psi(r_s,r_e)
\label{eq:SCHROEDINGER}
\end{equation}
Now, in the experiment, the beam of the traveling electron is split into
two sub-beams. Hence, the system's wave function can be written as a
sum of two terms
\begin{equation}
\Psi(r_s,r_e) = \phi _1(r_s)\psi _1(r_e) + \phi _2(r_s)\psi _2(r_e).
\label{eq:PSI}
\end{equation}
Here $\psi _i(r_e)$ is the traveling electron's wave function of the ith
sub beam and $\phi _i(r_s)$ is the corresponding wave function of the source.
Now, the traveling electron interacts with the charge at $r_s$ and
{\em vice versa}. For this reason $\phi _2$ {may} differ from $\phi _1$.
This analysis proves that {\em the phase is a property of a term and not
of a single particle}.
It is shown below how this result can help one to discern between correct
and incorrect claims of [33].

   Let us examine the magnetic AB effect. Here the source of the magnetic
field is a ring which is a single domain of a ferromagnetic material [34].
Thus, the source of the magnetic field is a quantum mechanical system. An
analysis of the interaction of a ferromagnetic atom with the field of
the traveling electron indicates that this interaction cannot induce
a quantum jump of an atom's state 
in the crystal [35,36]. Hence, in the case of the 
magnetic AB effect, the source can be treated as an inert object whose
state does not vary during the process.

   On the basis of this conclusion, one may cast the wave function
$(\!\!~\ref{eq:PSI})$ into the following form
\begin{equation}
\Psi(r_s,r_e) = \phi (r_s)[\psi _1(r_e) + \psi _2(r_e)],
\label{eq:INERT}
\end{equation}
where $\phi (r_s) = \phi _1(r_s) = \phi _2(r_s)$ denotes the inert state
of the ferromagnetic source. This outcome proves that, in the case of the
magnetic AB effect, $\phi (r_s)$ is factored out in $(\!\!~\ref{eq:INERT})$
and the phase of each term of the wave function
$(\!\!~\ref{eq:PSI})$ can be regarded as a single particle property. For
this reason, the magnetic AB's prediction is correct theoretically and
was detected in experiment [34].

   It was proved recently [35,36] that if the magnetic source is replaced 
by a classical device made of rotating charged 
material then the magnetic AB
effect disappears. The reason for this result is that the contribution
of the state of the traveling electron to the phase difference is
canceled by that of the (non-inert) source.

   The physics of the electric AB effect differs from that of
the magnetic one. Here the state
of the source {\em varies} during the process. A close examination of the
process proves that it is analogous to
the case of the classical magnet mentioned above. Thus, 
the contributions of the traveling electron and that of the source
to the phase difference cancel each other and the effect disappears
[37,38]. Moreover, if one adheres to
the AB's single particle approach [33,39], then energy conservation is
violated [37,38]. This outcome proves that the prediction of the
electric AB effect is wrong.

   The AB effects have a general (or philosophical) aspect too. Indeed,
in the AB processes, the traveling electron moves in a nonsimply
connected field free region. Thus, the single particle approach 
to the AB effects leads to the claim that topology is an inherent
element of quantum mechanics [33]. However, it can be proved that
this claim of AB has no profound meaning (see a detailed discussion
in [36], Section V). This conclusion can also be established on the
basis of the linearity of electrodynamics. Thus, the interaction $V$ is
a sum of 2-body interactions
\begin{equation}
V(r_s,r_e) = \sum _i V(r_{s_i} ,r_e),
\label{eq:V}
\end{equation}
where $r_{s_i}$ denotes the coordinates of the ith ferromagnetic atom. Here
no field free region exists because {\em the magnetic field of a single
ferromagnetic atom does not vanish at} $r_e$ and the magnetic field 
associated with the motion of the traveling electron does not vanish
at $r_{s_i}$. This analysis proves that the fundamental 2-body
interaction is {\em not} field free. Hence, the fundamental 2-body
interaction $(\!\!~\ref{eq:V})$ proves that 
the AB effects make no basis for
regarding the topological structure of field free regions as an inherent
property of quantum mechanics.

\vglue 0.66666in
\noindent
{\bf 7. Diffraction Free Beams}
\vglue 0.33333in

The idea that diffraction free beams (called also propagation invariant
beams) exist has been published in the literature [40]. The spatial
part of such a beam is assumed to take the form (see [40], eq. (2))
\begin{equation}
\phi = e^{i\beta z} J_0(a\rho)               
\label{eq:DF1}
\end{equation}
where $\rho$ denotes the radius in cylindrical coordinates, $J_0$
is the zeroth order Bessel function of the first kind and $a$ is a
factor having the dimension $[L^{-1}]$. Article [40]
has inspired a lot of activity and it has been cited more than 400
times.
Following [40], a family of diffraction free solutions of Maxwell
equations has been published [41].

Taking the diffraction free idea literally, one obviously realizes that
it is an error, because it is inconsistent with the uncertainty
principle. Indeed, the notion of a beam describes a set of physical
objects moving in a specific direction and the relevant cross section
containing these objects is much smaller than the beam's length (see
[40], p. 1499, near the bottom of the left column).

The ratio between the length and the diameter of the beam indicates that
it may be evaluated at the wave zone. It is easy to realize
that a Bessel beam like $(\!\!~\ref{eq:DF1})$
cannot exist [42]. Indeed, let us examine
a circle $C$ at the wave zone having a diameter which equals that
of the assumed beam (see fig. 3).
At the source, the beam's amplitude is a Bessel function, which 
means that it changes sign alternately. It follows that it interferes
{\em destructively} at $C$. Hence, since energy is conserved in the
process, one concludes that a part of the beam does not pass through $C$.
This conclusion means that the beam is {\em not} diffraction free.
Moreover, a Bessel beam spreads {\em faster} than a uniform beam because,
at circle $C$, interference of the latter is constructive.

Using this result, one infers that the family of diffraction free
solutions of Maxwell equations [41] describe solutions of electromagnetic
waves inside a perfect cylindrical wave guide.

Moreover, most (if not all)
experiments that follow [40] use a $\varphi$-invariant
setup and show a strong peak at the center. Now, the $\varphi$-invariant
solutions of of Maxwell equations [41] are derived from the following
vector potential
\begin{equation}
{\bf A} = -iJ_1(ar)e^{i(bz-\omega t)} {\bf u}_\varphi ,
\label{eq:A1}
\end{equation}
The fields are
\begin{equation}
{\bf E} = -\partial {\bf A}/\partial t
        = \omega J_1(ar)e^{i(bz-\omega t)} {\bf u}_\varphi .
\label{eq:E1}
\end{equation}
and
\begin{equation}
{\bf B} = curl {\bf A}
        = -b J_1(ar)e^{i(bz-\omega t)} {\bf u}_r -
              ia J_0(ar)e^{i(bz-\omega t)} {\bf u}_z .
\label{eq:B1}
\end{equation}
There is a dual solution where ${\bf E}\rightarrow \bf {B},\;
{\bf B}\rightarrow \bf {-E}$. Now, the Bessel function
$J_1(0) = 0$, which means that at the beam's 
center, energy current ${\bf E\times B}/4\pi $
of these solutions has a {\em minimum}. This
prediction contradicts the data and provides
another proof of the claim that the experiments should not be described
by diffraction free beams. A more detailed discussion of these topics can
be found in [42].

\vglue 0.66666in
\noindent
{\bf 8. Unexplained Quantum Chromodynamics Data}
\vglue 0.33333in

   The discussion presented in the previous Sections contain theoretical
arguments showing contradictions pertaining to several parts of
contemporary physics. This approach is analogous to an analysis of
errors in a mathematical theory. In addition to that, it is pointed out
in the introduction that a physical theory should satisfy a second kind
of tests - a compatibility of its predictions with experimental data.
Now, QCD is investigated for more than 30 years. Hence, one expects that
its main properties are already included in textbooks.
This Section contains a list of several experimental QCD data that have
no adequate explanation in textbooks.

\begin{itemize}
\item[{A.}]  The Higgs Mesons.

QCD assumes that quarks interact with 
particles called Higgs mesons. In spite
of a prolonged search, no evidence of these particles has been detected 
(see [4], p. 32).

\item[{B.}] The Photon-Hadron Interaction

The data show that a hard photon (having energy greater than 1000MeV)
interacts with a proton in a form which is very similar to that of a neutron
[28]. Due to the difference between the electric charge of 
proton's constituents 
and those of a neutron, this similarity cannot be explained
as interactions of the photon with electric charge. It turns out that
VMD (see Section 5) has been suggested in order to provide an
explanation for this effect. Now, it is proved in [30] that VMD contains
serious theoretical errors. Moreover, in 
the PACS classification it is regarded as just a model and in 
the xxx archive,
VMD is relegated to the phenomenological category. Hence, QCD has
no {\em theoretical} explanation for the interaction of a hard photon with
hadrons.

\item[{C.}] Properties of Anti-Quarks in Hadrons

The structure functions of proton constituents 
show that the width of $x$ values of antiquarks
is much smaller than that of quarks (see [43], p. 281). ($x$ is a
dimensionless Lorentz scalar used in the analysis.) Henceforth,
quarks and anti-quarks are denoted by $q$ and $\bar {q}$, respectively.
The width values indicate that, in the nucleon,
the uncertainty of momentum of $\bar {q}$
is smaller than the corresponding value of $q$ 
(see [43], pp. 270, 271). Therefore, due to the
uncertainty principle, one concludes that in a nucleon, $\bar {q}$
occupies a volume which is {\em larger} than that of $q$. This property of
nucleons lacks an adequate explanation.

In the literature, the $\bar {q}$ region is called "the 
$q-\bar {q}$ sea" (see [43] p. 281).
This terminology does not aim to be a theoretical explanation
and cannot be regarded as such.
Indeed, a $\pi $ meson is a bound state of $q\bar {q}$, both of
which came from the Dirac sea of negative energy states. Now, in
a $\pi $ meson, the $\bar {q}$  is attracted just by one $q$. In spite
of that, this force is strong enough for binding the system in a volume
which is even smaller than the nucleon's volume 
(see [4], pp. 499, 854). Hence, it is
not clear why 4 quarks (the 3 valence quarks and the $\bar {q}$'s
companion) cannot do that. It is concluded that QCD has no explanation
for the rather large volume of $\bar {q}$ in nucleons.

\item[{D.}] The Lack of Strongly Bound States of $qqqq\bar {q}$ 
(pentaquarks)

Consider the $qqqq\bar {q}$ 
system (a nucleon-meson system called pentaquark). The following
properties of hadrons is relevant to an evaluation of this object.
Data of strongly interacting systems show that gaps between energy states
are measured by hundreds of MeV. On the other hand, the binding energy
of a nucleon in a typical nucleus is about 8 MeV. These values can be used for
making a clear distinction
between true strong interactions and the nuclear force,
which is regarded as a residual force.

Another property of hadrons can be learnt from the data.
The mass of a $\pi$ meson is about 140MeV whereas the mass of a
nucleon is about 940MeV. Therefore, one concludes that
if QCD holds then the $q\bar {q}$
binding energy is much larger then that of a $qq$ pair (in a nucleon
there are 3 such pairs of interactions). 

Let us turn to the case of pentaquarks and examine
a particle called $\Theta ^+$ having a mass of 1540MeV.
Evidence of this object has been found in
several experiments (see e.g. [4], p. 916).
This object can be regarded as
a union of a neutron and a $K^+$ meson.
The sum of the masses of these particles is about 1435MeV. 
Therefore, the $\Theta ^+$ is an {\em unbound} state of the $nK^+$ system. 
On the other hand, a strongly bound state of $nK^+$ should have a mass
which is smaller than 1400MeV. Hence, QCD still does not provide an
explanation for the absence of {\em strongly} bound states of pentaquarks.
Moreover, it does not explain why the deuteron (a 6 quarks system) is
a bound state whereas the $nK^+$ (which contain an antiquark) has no
bound state.

\item[{E.}] The Uniform Density of Nuclear Matter

Consider nuclei that contain more than a very small number of nucleons. The
data show that for these nuclei, the nucleon density is (very nearly) the
same. QCD does not provide an explanation for these data. Another
aspect of this issue is that QCD does not provide an explanation for
the striking similarity between the form of the
van-der-Waals force and that of the nuclear force.

\item[{F.}] The EMC Effect

An examination of the mean volume occupied by quarks in nuclei
shows that it increases with the increase of the number of nucleons
of the nucleus [44,45]. This effect is analogous to the screening effect of
electrons in molecules. QCD has not predicted this effect and provides
no explanation for it.
\end{itemize}

In principle, one established experimental result which is inconsistent
with a theory, casts doubt on the theory's validity. In this Section one
can find several examples of experimental data which are not explained by QCD.

\vglue 0.66666in
\noindent
{\bf 9. Concluding Remarks}
\vglue 0.33333in

Two different aspects of the issues presented above are discussed in this
Section: implications of specific problems 
presented above and the general treatment of
theoretical errors by the community. These aspects are treated below
in this order.

The issues discussed above can be put in two different categories: 
issues having implications on other parts of theoretical physics
and stand alone topics.
It turns out that problems of
the Dirac monopole theory (see Section 2), those of
the VMD attempt to provide an
explanation for the hard photon - nucleon interaction (see Section 5)
and the experimental inconsistencies of QCD described in Section 8 are
related. Indeed, instead of the Dirac monopole theory, one can
construct a regular monopole theory [13,14]. It can be shown that this
monopole theory can explain 
experimental results which are unexplained by QCD [11].
Thus, the relations between the topics discussed in Sections 2,5 and 8
are probably the most significant part of this work.

   It is clear that there is a connection between the problems of  the
KG equation and those of the Yukawa theory, because these theories
examine the same kind of particle. The KG equation is supposed
to be the fundamental
equation of motion of a spin-0 particle whereas the Yukawa
theory examines this particle as an object that carries interaction
between two spin-1/2 particles. Hence, the difficulties of these
theories, which are presented in Sections 3 and 4, respectively,
have an underlying basis.

   On the other hand, the AB
effect and the Diffraction-Free idea can be regarded as 
stand alone issues. Thus, the
electric AB effect does not exist and the magnetic effect has no
inherent dependence on nonsimply connected field free regions of
space. Hence, one just concludes that the AB effects do not prove
that quantum mechanics has an inherent topological structure.

The Diffraction-Free idea is clearly inconsistent with the uncertainty
principle. Examining this idea literally, one concludes that it is just
wrong. Hence, fundamental physical theories are not affected by its
removal.

The general approach of a typical Journal of Physics to a free critical
debate of existing physical theories is very far from being
satisfactory. Indeed, a publication of Articles presenting pros and
cons concerning existing physical theories practically does not
exist in many Journals. One may wonder why the modern community of
physicists has adopted such a practice. After all, history of
scientific theories teaches us that
not all theories survive in the long run.
Another aspect of this matter is that the status of
a truly correct theory can only be improved if it is tested critically
every once in a while. Hence, people who genuinely believe in a
specific physical theory should support such a debate.

As a matter of fact, every topic presented in Sections 2-8 above cries
for a clarifying debate. A suppression of such a debate certainly does
not make a positive contribution to the progress of science. Referring
to this issue, it is interesting to cite S. D. Drell's final speech as
a president of the American Physical Society (APS). In his description of
referees of APS's Journals,
he uses the following quotation: ``We have met the enemy
and he is us" (see [46], p.61 second column). In my
personal experience, I have seen reports of many excellent referees.
However, there are too many referees belonging to a different category.
Considering them, I must say that I cannot deny Drell's description.


\newpage
References:
\begin{itemize}

\item[{[1]}] F. Rohrlich, {\em Classical Charged Particles},
(Addison-wesley, Reading mass, 1965). 
\item[{[2]}] P. A. M. Dirac, Proc. Royal Soc. {\bf A133}, 60 (1931).
\item[{[3]}] P. A. M. Dirac, Phys. Rev., {\bf {74}}, 817 (1948). 

\item[{[4]}] S. Eidelman et al. (Particle Data Group), Phys. Lett. {\bf B592}, 1 (2004).
\item[{[5]}] P. A. M. Dirac, A letter to A. Salam, published in 
{\em Monopoles in Quantum Field Theory}. Ed. N. S. Craigie, P. Goddard
and W. Nahm (World Scientific, Singapore, 1982).
\item[{[6]}] L. D. Landau and E. M. Lifshitz, {\em The Classical
Theory of Fields} (Pergamon, Oxford, 1975).
\item[{[7]}] J. D. Jackson, {\em Classical Electrodynamics} (John Wiley, New York,
1975). 
\item[{[8]}] D. Zwanziger, Phys. Rev., {\bf B137}, 647 (1965).
\item[{[9]}] S. Weinberg, Phys. Rev., {\bf B138}, 988 (1965).
\item[{[10]]}] C. R. Hagen, Phys. Rev., {\bf B140}, 804 (1965).
\item[{[11]}] E. Comay, published in {\em Has the Last Word Been Said on Classical Electrodynamics?} Editors: A Chubykalo, V Onoochin, A Espinoza, and
R Smirnov-Rueda (Rinton Press, Paramus, NJ, 2004). (The Article's
title is "A Regular Theory of Magnetic Monopoles and Its Implications".)
\item[{[12]}] P. Goddard and D. I. Olive, Rep. Prog. Phys.,
{\bf 41}, 1357 (1978).
\item[{[13]}]  E. Comay, Nuovo Cimento, {\bf 80B}, 159 (1984).
\item[{[14]}] E. Comay, Nuovo Cimento, {\bf 110B}, 1347 (1995).
\item[{[15]}] E. Comay, Lett. Nuovo Cimento {\bf 43}, 150 (1985).
\item[{[16]}] H. Feshbach and F. Villars, Rev. Mod. Phys. 
{\bf 30} 24-45 (1958).
\item[{[17]}] W. Pauli and V. Weisskopf, Helv. Phys. Acta, {\bf 7}, 709 (1934).
\item[{[18]}] 
An English translation of [17] can be found in A. I. Miller {\em Early Quantum
Electrodynamics} (University Press, Cambridge, 1994). pp. 188-205.
\item[{[19]}] P. A. M. Dirac, {\em Mathematical Foundations of Quantum
Theory} Ed. A. R. Marlow (Academic, New York, 1978). (See pp. 3,4). 
\item[{[20]}] Weinberg S 1995 The Quantum Theory of Fields
(Cambridge: University Press). Vol. 1. 
\item[{[21]}] E. Comay, Apeiron {\bf 12}, no. 1, 27 (2005).
\item[{[22]}] E. Comay, Apeiron, {\bf 11}, No. 3, 1 (2004).
\item[{[23]}]  L. D. Landau and E. M. Lifshitz, {\em Quantum Mechanics}
(Pergamon, London, 1959). P. 6.
\item[{[24]}] M. E. Peshkin and D. V. Schroeder, {\em An Introduction to 
Quantum Field Theory} (Addison-Wesley, Reading, Mass., 1995).
\item[{[25]}] G. Sterman {\em An Introduction to Quantum Field Theory}
(University Press, Cambridge, 1993). 
\item[{[26]}] J. D. Bjorken and S. D. Drell {\em Relativistic Quantum
Mechanics} (McGraw, New York, 1964).
\item[{[27]}] S. S. M. Wong, {\em Introductory Nuclear Physics} (Wiley, 
New York, 1998). 2nd edition.
\item[{[28]}] T. H. Bauer, R. D. Spital, D. R. Yennie and F. M. Pipkin,
Rev. Mod. Phys. {\bf 50}, 261 (1978).
\item[{[29]}] H. Frauenfelder and E. M. Henley, {\em Subatomic
Physics}, (Prentice Hall, Englewood Cliffs, 1991). pp. 296-304.
\item[{[30]}] E. Comay, Apeiron {\bf 10}, no. 2, 87 (2003).
\item[{[31]}] E. P. Wigner, Annals of Math., {\bf 40},
149 (1939). 
\item[{[32]}] S. S. Schweber, {\em An Introduction to Relativistic
Quantum Field Theory}, Harper \& Row, New York, 1964. pp. 44-53.
\item[{[33]}] Y. Aharonov and D. Bohm Phys. Rev. {\bf 115} 485 (1959).
\item[{[34]}] A. Tonomura et al., Phys. Rev. Lett. {\bf 56},
792 (1986).
\item[{[35]}] E. Comay, Phys. Lett {\bf A250}, 12 (1998).
\item[{[36]}] E. Comay, Phys. Rev. {\bf A62}, 042102-1 (2000).
\item[{[37]}] E. Comay, Phys. Lett {\bf A120} 196 (1987).
\item[{[38]}] E. Comay, Phys. Lett {\bf A125} 403 (1987).
\item[{[39]}] Y. Aharonov and D. Bohm Phys. Rev. {\bf 123} 1511 (1961).
\item[{[40]}] J. Durnin, J. J. Miceli, Jr. and J. H. Eberly,
Phys. Rev. Lett. {\bf 58}, 1499 (1987).
\item[{[41]}] Z. Bouchal and M. Olivik, J. Mod. Opt.
{\bf 42}, 1555 (1995).
\item[{[42]}] E. Comay, {\em Focus on Lasers and Electro-Optics Research}
Ed: W. T. Arkin, p. 273.
\item[{[43]}] D. H. Perkins, {\em Introduction to High Energy Physics}
(Addison-Wesley, Menlo Park, CA, 1987). 
\item[{[44]}] J. J. Aubert et al. (EMC), Phys. Lett. {\bf 123B},
275 (1983).
\item[{[45]}] A. Bodek et al., Phys. Rev. Lett. {\bf 50}, 1431 (1983).

\item[{[46]}] S. D. Drell, Physics Today, {\bf 40}, 56 (August 1987).
\end{itemize}

\newpage
\noindent
Figure Captions

\noindent
Figure 1: 

   Two rays of light are emitted from sources $S_1$ and $S_2$
which are located at $x=\pm 1$, respectively. The
rays intersect at point $O$ which is embedded in the $(x,y)$ plane.
(This figure is published in [29] and is used here with permission.)

\noindent
Fig. 2:

A beam consists of electrons that travel from left to right .
They are split into two
sub-beams at point $A$. The sub-beams travel in a field free region and
interfere on the screen $S$. The field is nonzero in a region $R$ denoted by
the black circle.

\noindent
Fig. 3: \\

A beam of electromagnetic wave is emitted from a 
circular source $S$. The beam's intensity is calculated at a circle $C$
whose radius is the same as that of the source. (This figure was
published in [41]).

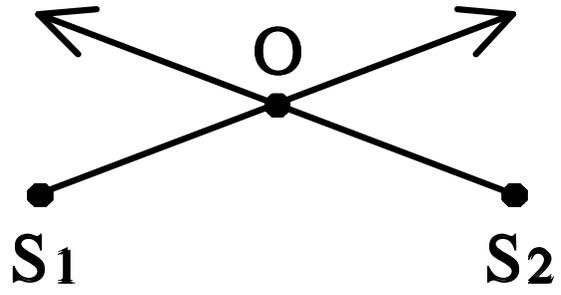
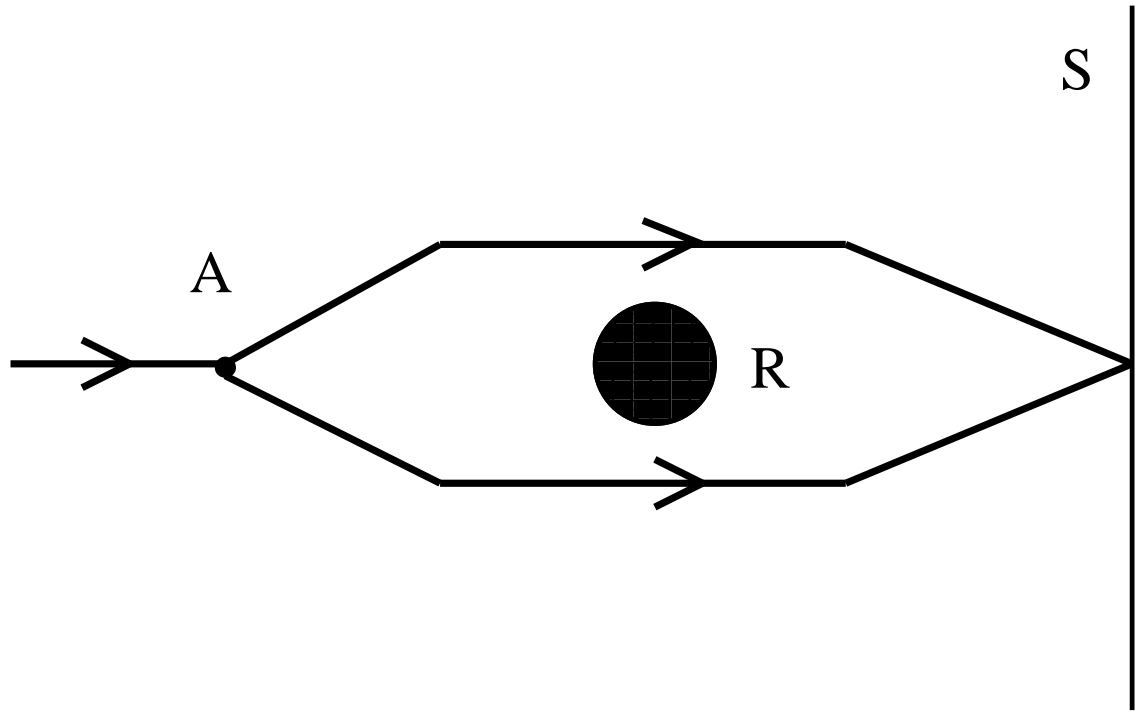
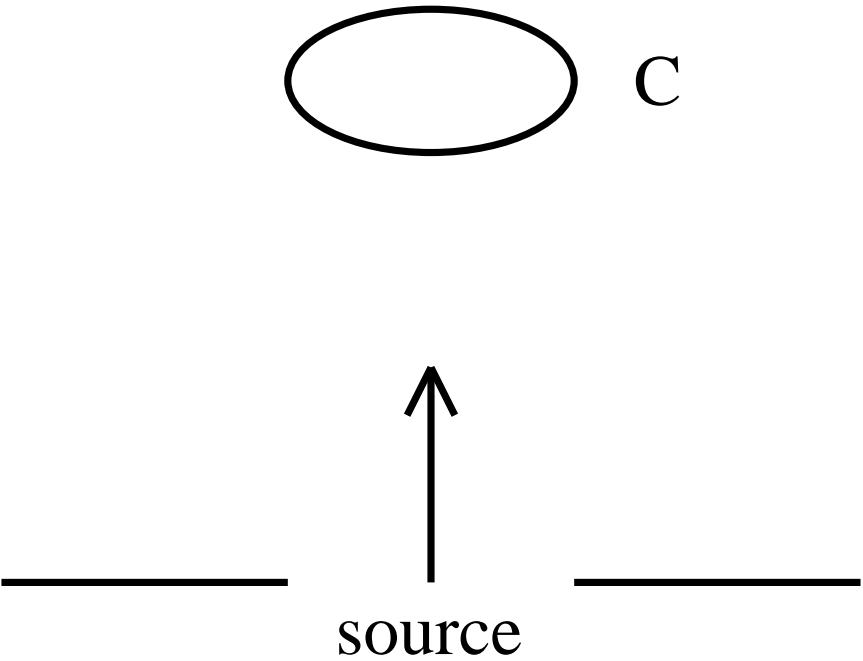
\end{document}